\documentclass{aastex}
\usepackage{emulateapj5}

\begin{document}

\title{The DIRTY Model. I. Monte Carlo Radiative Transfer Through Dust}

\author{Karl D.\ Gordon\altaffilmark{1},
   K.\ A.\ Misselt\altaffilmark{2,3,4},
   Adolf N.\ Witt\altaffilmark{5},
   and Geoffrey C.\ Clayton\altaffilmark{2}}
\altaffiltext{1}{Steward Observatory, University of Arizona,
   Tucson, AZ 85721; kgordon@as.arizona.edu}
\altaffiltext{2}{Department of Physics \& Astronomy, Louisiana State
   University, Baton Rouge, LA 70803; gclayton@fenway.phys.lsu.edu}
\altaffiltext{3}{present address: NASA/Goddard Space Flight Center,
   Code 685, Greenbelt, MD 20771; misselt@idlastro.gsfc.nasa.gov}
\altaffiltext{4}{National Research Council/Resident Research Associate}
\altaffiltext{5}{Ritter Astrophysical Research Center, The University
   of Toledo, Toledo, OH 43606; awitt@dusty.astro.utoledo.edu}

\shortauthors{Gordon et al.}
\shorttitle{DIRTY I}

\begin{abstract}
We present the DIRTY radiative transfer model in this paper and a
companion paper.  This model computes the polarized radiative transfer
of photons from arbitrary distributions of stars through arbitrary
distributions of dust using Monte Carlo techniques.  The dust
re-emission is done self-consistently with the dust absorption and
scattering and includes all three important emission paths:
equilibrium thermal emission, non-equilibrium thermal emission, and
the aromatic features emission.  The algorithm used for the radiative
transfer allows for the efficient computation of the appearance of a
model system as seen from any viewing direction.  We present a simple
method for computing an upper limit on the output quantity
uncertainties for Monte Carlo radiative transfer models which use the
weighted photon approach.
\end{abstract}


\section{Introduction}

Dust radiative transfer models are crucial to interpreting many
astronomical observations ranging from reflection nebulae with a
single star to galaxies comprised of stars, gas, and dust arranged in
complex geometries.  Almost all observations are affected by the
presence of dust intrinsic to the system under study or present in the
foreground.  For dust present in the foreground as is the case for
observations of single stars, the effects of dust can be accurately
taken into account using a screen geometry.  In this geometry, the
effects are well characterized by an extinction curve which is
determined by the amount and intrinsic properties (e.g., sizes and
materials) of the dust \citep{cla00}.  However, when the dust is near
or mixed with the system being studied, its effects become dependent
not only on the intrinsic properties of the dust grains, but on the
geometry of the stars and dust in the system as well \citep{wit92,
wit96, wit00}.  Since the effects of mixing stars and dust are
non-trivial, a radiative transfer model is required.  Using such a
model produces a deeper understanding of the physical properties of
the system being studied.  This applies to a wide variety of
astrophysical systems, from reflection nebulae to galaxies.  There are
a variety of numerical methods other than Monte Carlo techniques which
can be used to solve the radiative transfer through dust.  One example
is the doubling method \citep{zub00}.  While the other methods can
achieve faster computations for specific cases, Monte Carlo techniques
are the most flexible and allow the study of a wider range of
astrophysical systems with arbitrary geometries.

There exist many Monte Carlo dust radiative transfer models which have
been constructed to study a variety of problems.  In most models, the
computational needs of the Monte Carlo technique have been reduced by
exploiting symmetries and/or making approximations.  This necessarily
restricts the variety of systems to which a particular model can be
successfully applied.  Reflection nebulae and circumstellar shells
have been successfully approximated as spherically and azimuthally
symmetric systems \citep{wit77, wit82, vos94, cal95, wit96}.  Models
for bipolar nebulae, associated with young stellar objects and active
galactic nuclei, usually exploit the azimuthal symmetry of such
systems \citep{yus84, whi92, fis94, wod98, you00}.  The general
behavior of dust in galaxies has been studied using the concept of
spherical galactic environments \citep{wit92, wit00}.  Spiral galaxies
have been studied with models using the approximation of disk galaxies
\citep{bia96, wod97, kuc98}.

The importance of the multi-phase nature of dust distributions has
motivated the construction of models which can handle arbitrary
distributions of dust.  Models which can handle such dust
distributions include those of \citet{wol98}, \citet{var99},
\citet{wod99}, \citet{bia00}, and our model.  More recently, the
importance of the re-emission of the energy absorbed by dust has been
acknowledged and increasingly added to some radiative transfer models.
The dust re-emission can be broken into three components; thermal
equilibrium emission (large particles), non-equilibrium thermal
emission (small particles), and the aromatic features (PAH-like
particle emission between 3.3 and $\sim$20~$\micron$).  The inclusion
of the dust re-emission to radiative transfer models has been done
with a range of approximations.  The easiest approximation is to
include dust in thermal equilibrium \citep{var99, wol99, bia00b}.
Including the non-equilibrium emission and aromatic feature emission
is much harder and, as a result, requires a significant amount of CPU
time.  Due to the complexity of the non-equilibrium and aromatic
feature emission, some models use a simplified \citep{sil98} and/or
empirical approach to including them.  The model described in this
paper and a companion paper \citep{mis00} can handle arbitrary dust
distributions and fully includes all three dust emission components
self-consistently.

The DIRTY model (DustI Radiative Transfer, Yeah!) grew out of
spherically symmetric models which were used to study reflection
nebulae \citep{wit82,gor94,cal95}.  Using Monte Carlo techniques,
these models computed the radiative transfer through spherical shells
of dust illuminated by point sources.  The model used by
\citet{gor94} allowed for multiple illuminating stars arbitrarily
distributed throughout a spherical nebula and the images of such a
system could be constructed for arbitrary lines-of-sight.  As such,
the \citet{gor94} model is the direct ancestor of the DIRTY model.  It
is important to note that the algorithm used by \citet{gor94}, while
giving the correct total scattered light value, gave too flat a
distribution of scattered light across the system.  The correct
algorithm for efficient computation of the appearance of a system for
a particular line-of-sight is described by \citet{yus84}.

The motivation for the DIRTY model grew out of the conference ``Dust
Survival in Interstellar/Intergalactic Media'' held at the Space
Telescope Science Institute in 1994.  One of the major concerns at
this conference was that all Monte Carlo radiative transfer models at
that time assumed a smooth distribution of dust, yet the interstellar
medium is known to be clumpy over a large range of size scales
\citep{col80, sca90, ros95}.  We constructed the DIRTY model to answer
this concern.  The main change was to move from spherical shells to
rectangular cells as the basic unit of the dust density distribution.
Using the simple system of a spherical nebula with a central
illuminating star, \citet{wit96} explored the differences between the
radiative transfer in smooth and clumpy 3-dimensional dust
distributions.  The implications of clumpy dust for galaxies were
discussed by
\citet{wit00} using the concept of spherical galactic environments.
The DIRTY model has been applied to starburst galaxies \citep{gor97,
gor00}, spiral galaxies \citep{kuc98}, the UW Cen reflection nebula
\citep{cla99}, and the nucleus of M33 \citep{gor99}.

The DIRTY model was constructed to compute the radiative transfer of
photons from arbitrary distributions of photon emitters through
arbitrary distributions of dust.  This model self-consistently models
the scattering and absorption of photons through dust and the
re-emission of photons from dust.  The algorithm for the radiative
transfer, including polarization, is based on the work of
\citet{wit77}, \citet{yus84}, and \citet{cod95}.  The details of this
algorithm are the subject of this paper.  The re-emission of energy
absorbed by the dust is the subject of a companion paper
\citep{mis00}.  The radiative transfer of the re-emitted photons is
handled by the same algorithm as the stellar photons.  This requires
an iterative procedure to account for dust absorption and scattering
of re-emitted photons and their subsequent re-emission.  The details
of the algorithms for the dust emission and iterative procedure are
given by \citet{mis00}.

\section{Monte Carlo Radiative Transfer}

The goal of the DIRTY model is to compute the radiative transfer of
photons through arbitrary distributions of dust.  The lack of
symmetries motivated us to use Monte Carlo techniques which are based
on the work of \citet{wit77}, \citet{yus84}, and \citet{cod95}.  The
forward scattering nature of dust grains \citep{gor94, gor97} and the
observed multi-phase characteristics of the distribution of dust
strongly argue for radiative transfer models based on Monte Carlo
techniques.  Such techniques rely on the probabilistic understanding
of the interaction of a photon with a dust grain and sufficient
computing power to evaluate this interaction for many paths through
the distribution of the dust.  This approach allows us to efficiently
compute what a system of stars and dust will look like for any
line-of-sight.  In addition to the algorithm described in
\S\ref{sec_algorithm}-\ref{sec_unc}, the model requires the physical
description of the dust grain properties, the dust distribution, the
distribution of photon emitters, and a pseudo-random number generator.

The physical description of the dust grain properties is given by the
wavelength dependent behavior of the dust optical depth, albedo, and
scattering phase function.  The optical depth determines the
probability of a photon interaction with a dust grain, the albedo
determines the probability of the dust grain scattering (or absorbing)
the photon, and the scattering phase function gives the probability of
the photon being scattered at a specific angle as well as the change
in the polarization state of the photon as a result of the scattering.
We have taken the physical description of the dust properties from
work by \citet{cla00} and \citet{kim94}.  \citet{cla00} used MEM
techniques to model the size distribution of dust grains for dust
extinction curves in the Milky Way, Large Magellanic Cloud, and Small
Magellanic Cloud.  This work provides albedos and scattering phase
functions appropriate for the range of known interstellar dust
extinction curves.

The DIRTY model allows for arbitrary distributions of dust limited
only by the amount of memory available to store the distribution.
This is accomplished by representing the dust distribution with a
3-dimensional grid.  Each grid cell represents a region of uniform
dust density.  Thus, any dust geometry can be represented with the
smallest scale of inhomogeneity being the size of a grid cell.  The
physical dimensions of each axis of the grid are specified by separate
1-dimensional arrays.  This allows the physical dimensions of one axis
to vary making it possible to efficiently model both cubical
geometries as well as thin pancake-like geometries.  In addition, this
geometry allows for the representation of multi-phase or clumpy dust
distributions.  In the simplest form, this is a two-phase medium with
high-density clumps and a low density inter-clump region.  This type
of dust distribution can be described using three parameters: the
fractional filling factor of the high density dust ($ff$), the density
ratio between the low and high density dust ($k_2/k_1$), and the size
of a grid cell in comparison to the system size ($1/N$, where $N$ is
the number of grid cells on an axis).  The differences in the
radiative transfer between homogeneous and two-phase dust
distributions is discussed at length by \citet{wit96}.

The tracking of the photons through the 3-dimensional grid is simple
and quite efficient.  A photon's location and direction of travel is
uniquely defined by its x, y, and z positions and the $u$, $v$, and
$w$ direction cosines.  When a photon enters a grid cell, the side
through which it exits can be quickly computed in the following
manner.  Using the x, y, and z dimensions of the cell and the
direction cosines of the photon, the distance the photon would have to
travel in each of the x, y, and z directions to exit the cell through
a plane parallel to the wall of the cell in those directions can be
computed.  The photon exits through the side which results in the
shortest distance traveled.  The photon's position can be stepped by
this distance and ancillary information updated (e.g., optical depth
traveled).

Just as our model permits the use of arbitrary dust distributions, it
also allows photons to be emitted from arbitrary source distributions.
These can be a single star, a distribution of stars, or any definable
surface (e.g., an accretion disk).  The sources can emit isotropically
or only in specific directions.  We use the ``ran2'' pseudo-random
number generator which is described by \citet{pre92} and has a period
greater than $2
\times 10^{18}$. 

\subsection{Radiative Transfer Algorithm - A Photon's Life
    \label{sec_algorithm}}

The transfer of photons through dust using Monte Carlo techniques is
historically based on following a single photon through a distribution
of dust.  In this paradigm, the chance of the photon being absorbed or
scattered is determined using a random number generator with the
results dictating the life of that photon.  Therefore, a particular
photon might exit the system without interacting with a dust grain, be
absorbed on its first scattering (or subsequent scatterings), or
scatter one or more times before exiting the system.  This is not a
particularly efficient method of computing the direct, scattered, or
absorbed light from a system as each photon can contribute to only one
of these quantities.  In addition, each photon exits the system in a
particular direction which is not necessarily the direction from which
one wants to observe the system.  The efficiency of the Monte Carlo
method can be increased greatly by assigning the photon a weight which
allows every photon to contribute to the absorbed energy, the direct
light, and the scattered light for a particular line-of-sight
\citep{wit77, yus84}.

The algorithm we use for the radiative transfer in the DIRTY model is
based on the work of \citet{wit77}, \citet{yus84}, and \citet{cod95}.
In the following description of our algorithm, we will use the concept
of the life of a photon.  To track each photon's state, we tabulate
its position, direction of travel (via direction cosines $u$, $v$, and
$w$), weight, and polarization (via Stokes I, Q, U, and V values).

A photon is born according to an input source distribution and its
initial direction is usually chosen from an isotropic distribution
(e.g., eq.\ 5-7 of \citet{wit77}).  The assumption of isotropic
emission can be modified according to the particular problem being
studied.  For example, a mask with different sized holes at specific
locations was applied in the DIRTY model to the central star of the
nebula surrounding the R Corona Borealis star UW Cen
\citep{cla99}.  In this case, only the photons emitted in the
direction of the holes in the mask were allowed to move to the next
step.  The initial weight of the photon is usually
\begin{equation}
W_0^\alpha = L^\alpha,
\end{equation}
where $L^\alpha$ is the luminosity associated with the $\alpha$th
photon and $\alpha$ runs from 1 to $N$.  Usually $L^\alpha = L/N$
where $L$ is the luminosity of the source distribution at the modeled
wavelength and $N$ is the number of photons in the model run.  The
value of $L^\alpha$ can be varied depending on the particular problem
being studied to optimize the model running time \citep{yus84}.  The
photon is initially assumed to be unpolarized (i.e. $S_0^\alpha =
(I_0^\alpha, Q_0^\alpha, U_0^\alpha, V_0^\alpha) = (1, 0, 0, 0)$).

The flux corresponding to the part of the photon which escapes from
the system in the direction of the observer is
\begin{equation}
F_0^\alpha = W_0^\alpha e^{-\tau(obs)_0^\alpha} \frac{1}{4\pi d^2},
\label{eq_fd}
\end{equation}
where $\tau(obs)_0^\alpha$ is the optical depth along the path from
the birth position of the photon to the surface of the system in the
direction of the observer and $d$ is the distance to the system being
modeled.  In general, $\tau(obs)_0^\alpha$ is different for each
photon emitted unless the system being modeled has a single central
point source embedded in a spherical, homogeneous dust distribution.

The first scattering of the photon is forced to insure that every
photon contributes to the scattered light \citep{cas59}.  The optical
depth to the first scattering site is
\begin{equation}
\tau_1^\alpha = -\ln \left[ 1 - \xi 
    \left( 1 - e^{-\tau_s^\alpha} \right) \right]
\end{equation}
where $\xi$ is a random number between 0 and 1 and $\tau_s^\alpha$ is
the optical depth to the surface of the nebula in the direction the
$\alpha$th photon is traveling.  The weight of the photon after the
first scattering is then
\begin{equation}
W_1^\alpha = W_0^\alpha a \left( 1 - e^{-\tau_s^\alpha} \right)
\end{equation}
where $a$ is the dust albedo.  The fraction of the photon which is
absorbed at the scattering site is
\begin{equation}
A_1^\alpha = W_0^\alpha (1 - a) \left( 1 - e^{-\tau_s^\alpha} \right).
\label{eq_a1}
\end{equation}
The flux corresponding to the fraction of the photon which is
scattered at the first scattering toward the observer is
\begin{equation}
F_1^\alpha = W_1^\alpha e^{-\tau(obs)_1^\alpha} \Phi
(\theta(obs)_1^\alpha) \frac{1}{4\pi d^2},
\label{eq_fs1}
\end{equation}
where $\tau(obs)_1^\alpha$ is the optical depth from the first
scattering site along the direction towards the observer,
$\Phi(\theta)$ is the scattering phase function, and
$\theta(obs)_1^\alpha$ is the angle between the direction the photon
was traveling before the scattering and the direction towards the
observer.  This angle is easily calculated for the $n$th scattering
using
\begin{equation}
\cos (\theta(obs)_n^\alpha) = (u_{n-1}^\alpha u_{obs} + 
	v_{n-1}^\alpha v_{obs} + w_{n-1}^\alpha w_{obs})
\end{equation}
where $u_{n-1}$, $v_{n-1}$, and $w_{n-1}$ are the direction cosines of
the photon before it scatters for the $n$th time and $u_{obs}$,
$v_{obs}$, $w_{obs}$, are the direction cosines from the $n$th
scattering site towards the observer.  This equation is equivalent to
eq.~17 of \citet{yus84}.

The polarization state of the flux corresponding to the fraction of
the photon which reaches the observer from the $n$th scattering is
calculated using
\begin{equation}
\label{eq_new_pol}
S_n^\alpha = L(\pi - i_2^\alpha)
R(\theta(obs)_n^\alpha)L(-i_1^\alpha)S_{n-1}^\alpha
\label{eq_sn}
\end{equation}
where $S_n^\alpha$ is the new Stokes vector, $L(-i_1^\alpha)$ rotates
$S_{n-1}^\alpha$ from the reference frame to the scattering plane,
$R(\theta)$ is the scattering matrix, and $L(\pi - i_2^\alpha)$
rotates the new Stokes vector from the scattering plane to the
reference frame.  The reference frame for the Stokes vector is set to
be the z-axis.  The rotation matrix is
\begin{equation}
L(\Psi) = \left[ \begin{array}{cccc} 1 & 0 & 0 & 0 \\ 0 & \cos (2\Psi)
& \sin (2\Psi) & 0 \\ 0 & - \sin (2\Psi) & \cos (2\Psi) & 0 \\ 0 & 0 &
0 & 1 \\ \end{array} \right]
\end{equation}
where $\Psi$ is the rotation angle.  The last two equations are eqs.\
2-3 of \citet{cod95}.

The scattering matrix, $R(\theta)$, is the $4\times 4$ Mueller matrix
and is also known as the phase matrix \citep{boh83, mish00}.  For this
work, we have used calculations of $R(\theta)$ by \citet{cla00} for
size distributions of spherical dust grains.  As a result, there are
only 8 nonzero elements of $R(\theta)$, with $R_{11}=R_{22}$,
$R_{12}=R_{21}$, $R_{33}=R_{44}$, and $R_{34}=-R_{43}$.  The angles
$i_1^\alpha$ and $i_2^\alpha$ are calculated from the knowledge of the
direction the photon is traveling and $\theta(obs)_n^\alpha$ using
spherical geometry.  A useful figure for visualizing the calculation
of $i_1^\alpha$ and $i_2^\alpha$ is Fig.~1 of \citet{cod95}.  The
Stokes polarization vector is renormalized after each scattering so
that $I_n^\alpha = 1$.  As a result, the flux contained in the Q
polarization component is just $Q_n^\alpha F_n^\alpha$.  Similar
equations give the U and V polarization component fluxes.

The direction into which the photon is scattered at the $n$th
scattering is described by the spherical angles $\theta_n^\alpha$ and
$\phi_n^\alpha$.  The $\theta_n^\alpha$ angle is determined from the
the dust scattering phase function.  We use either a Henyey-Greenstein
phase function or the full scattering matrices.  The Henyey-Greenstein
phase function is
\begin{equation}
\Phi(\cos(\theta),g) = \frac{1 - g^2}
   {4\pi \left[ 1 + g^2 - 2g\cos(\theta) \right]^{3/2}}
\end{equation} 
which is eq.~3 of \citet{wit77} and where $g = \left< \cos \theta
\right>$ is the scattering phase function asymmetry.
For the Henyey-Greenstein phase function, the cosine of the scattering
angle is calculated using
\begin{equation}
\cos \theta_n^\alpha = \frac{1}{2g} \left[ (1 + g^2) 
  - \left( \frac{1 - g^2}{1 - g + 2g\xi} \right)^2 \right]
\label{eq_hg_invert}
\end{equation}
which is eq.\ 19 of
\citet{wit77}.  For the full scattering matrices, the scattering angle
is determined from numerically inverting the function
\begin{equation}
\Phi(\theta_n^\alpha) = I_{n-1}^\alpha R_{11}(\theta_n^\alpha) +
  Q_{n-1}^\alpha R_{22}(\theta_n^\alpha),
\label{eq_scat}
\end{equation}
where $R_{11}(\theta_n^\alpha)$ and $R_{22}(\theta_n^\alpha)$ are
elements of the scattering matrix and $I_{n-1}^\alpha$ and
$Q_{n-1}^\alpha$ are elements of the photon's Stokes vector.  Unlike
\citet{cod95} who approximate $\Phi(\theta)$ as $R_{11}(\theta)$, we
use the polarization dependent scattering phase function.  The
$\phi_n^\alpha$ angle of scattering is determined using
\begin{equation}
\phi_n^\alpha = \pi \left(2\xi - 1\right).
\label{eq_phi}
\end{equation}
Given the direction into which the photon is scattered
($\theta_n^\alpha$ and $\phi_n^\alpha$), the new direction cosines can
be computed from the old direction cosines (e.g., eq.\ 22 of
\citet{wit77}).  The polarization state of the photon after the $n$th
scattering is computed using eq.~\ref{eq_new_pol} with
$\theta_n^\alpha$ replacing $\theta(obs)_n^\alpha$.

The subsequent scatterings are not forced.  By forcing the first
scattering we insure that every photon contributes to the scattered
flux.  By not forcing the subsequent scatterings we insure that the
calculation is only done for the scatterings which contribute
significantly to the scattered flux.  The optical depth to the $n$th
scattering, $\tau_n^\alpha$, is determined using
\begin{equation}
\tau_n^\alpha = - \ln \xi
\end{equation}
which holds for $n \geq 2$.  If $\tau_n^\alpha$ is greater than the
optical depth to the surface in the direction the photon is traveling,
the photon escapes the system.  If not, the photon is scattered in a
direction determined using eqs.~\ref{eq_hg_invert}-\ref{eq_phi}.  The
weight of the photon after the $n$th scattering is
\begin{equation}
W_n^\alpha = a W_{n-1}^\alpha
\end{equation}
which holds for $n \geq 2$.  The fraction of the photon which is
absorbed at the scattering site is
\begin{equation}
A_n^\alpha = (1 - a) W_{n-1}^\alpha
\label{eq_an}
\end{equation}
which holds for $n \geq 2$.  The fraction of the photon which is
scattered towards the observer is
\begin{equation}
F_n^\alpha = W_n^\alpha e^{-\tau(obs)_n^\alpha} \Phi
(\theta(obs)_n^\alpha) \frac{1}{4\pi d^2}
\label{eq_fsn}
\end{equation}
which holds for $n \geq 1$.  The total number of scatterings a photon
undergoes is limited to be less than $n_{max}$.  For low to moderate
optical depths, we set to $n_{max} = 50$ to avoid spending significant
computation time on photons whose weight is very small as $W_n^\alpha
\propto a^n$.  It is important to remember that there are times (e.g.,
high $\tau$ values) when the value of $n_{max}$ should be set to a
higher value.  For example, if it is important to know the energy
absorbed in very high optical depth parts of a dust distribution the
value of $n_{max}$ should be increased.

\subsection{Integrated Quantities}

While the previous section has described in detail the life of a
single photon, the results we are interested in are integrated
quantities describing the system being modeled.  These integrated
quantities are images of the system (e.g., direct and scattered light
images) and the 3-dimensional matrix of the energy which was absorbed
by the dust.

Images of the system are constructed by projecting the birth position
(for the direct light image) or the $n$th scattering position (for the
scattered light image) of the photon onto the sky.  In the DIRTY
model, we embed the dust density grid in a sphere making the sky a
plane which is tangent to the sphere in the direction of the observer.
This plane is divided into rectangular pixels and the images are built
up as each photon is run through the model.  Special care must be
applied when constructing the images which give the polarization state
of the scattered flux, specifically the Q and U images.  For the
radiative transfer of the photon, the polarization state has been
referenced to the z-axis of the system.  As part of the construction
of these images, the polarization vector of the photon must be rotated
so that it is referenced to the y-axis of the image plane.

The direct light surface brightness image is constructed using
\begin{equation}
I_D(i,j) = \frac{1}{\Omega_{pixel}} \sum_{\alpha} F_0^\alpha
\label{eq_idij}
\end{equation}
where $F_0^\alpha$ is the direct flux for the $\alpha$th photon (see
eq.~\ref{eq_fd}), $\Omega_{pixel}$ is the surface area of pixel
$(i,j)$ in steradians, and the sum over $\alpha$ is done only for
photons whose birth position fall within pixel $(i,j)$ when projected
onto the sky.

The scattered light surface brightness image is constructed using
\begin{equation}
I_S(i,j) = \frac{1}{\Omega_{pixel}} \sum_{\alpha} \sum_n F_n^\alpha
\label{eq_is_image}
\end{equation}
where $F_n^\alpha$ is the scattered flux for the $n$th scattering of
the $\alpha$th photon (see eqs.~\ref{eq_fs1} and \ref{eq_fsn}), and
the sums over $\alpha$ and $n$ are done only for photons and
scatterings of that photon whose scattering sites fall within pixel
$(i,j)$ when projected onto the sky.  The images which describe the
polarization state of the scattered light image are constructed by
first rotating the $S_n^\alpha$ Stokes vector from the coordinated
system referenced to the z-axis to a coordinate system referenced to
the y-axis of the image plane.  This is done using
\begin{equation}
{S'}_n^\alpha = L(\beta^\alpha)S_n^\alpha
\end{equation}
where $\beta^\alpha$ is the angle which rotates the Stokes vector
between the two coordinate systems.  The Q image, and similarly the U
and V images, are constructed using
\begin{equation}
Q(i,j) = \frac{1}{\Omega_{pixel}} \sum_{\alpha} \sum_n {Q'}_n^\alpha
F_n^\alpha
\label{eq_qij}
\end{equation}
where the sums over $\alpha$ and $n$ are done over the same limits as
eq.~\ref{eq_is_image}.  Of course, the sum of the direct and scattered
light images gives the image one would observe at a telescope.

The 3-dimensional absorbed energy matrix is calculated using
\begin{equation}
A(i,j,k) = \sum_{\alpha} \sum_n A_n^\alpha
\end{equation}
where the sum over $\alpha$ and $n$ are only done over scatterings
which happen in cell $(i,j,k)$ and $A_n^\alpha$ is the energy which is
absorbed the $n$th scattering site of the $\alpha$th photon (see
eqs.~\ref{eq_a1} and \ref{eq_an}).  The 3D absorbed energy matrix is
what is needed to compute the dust emission spectrum \citep{mis00}.

\subsection{Uncertainties \label{sec_unc}}

Due to our use of photon weights, the uncertainties in output
quantities (i.e., scattered intensity, polarization, etc.) can not be
computed directly from the square root of the number of photons as is
usually done when non-weighted Monte Carlo techniques are used.  The
ability to calculate uncertainties is crucial when using Monte Carlo
techniques as the accuracy of model results are dependent on the
number of photons run.  If such uncertainties can be calculated during
the model run, they can be used to dynamically set the number of
photons needed in a model run to achieve a user input accuracy.  We
have adopted a simple technique for computing the upper limits on the
uncertainties in the output quantities.  If the output quantity is
$X$, then
\begin{equation}
X = \sum_{\alpha=1}^N \sum_n x_n^\alpha = M\overline{x}
\end{equation}
where $x_n^\alpha$ is the contribution of the $n$th scattering of the
$\alpha$th photon to $X$, $M$ is the total number of number photons or
scatterings in the model run, and $\bar{x}$ is the average
contribution each photon or scattering makes to $X$.  In the case of
the direct flux, the sum over $n$ is dropped and $M = N$.  The
uncertainty in $X$ is then
\begin{equation}
\sigma_X = X \frac{\sigma_x}{\overline{x}}
\end{equation}
where $\sigma_x$ is the standard deviation of $\bar{x}$.  The value of
$\sigma_x$ is calculated using
\begin{eqnarray}
\sigma_x^2 & = & \frac{1}{M(M-1)}\sum_{\alpha=1}^N \sum_n
   (x_n^\alpha - \overline{x})^2 \nonumber\\ & = & \frac{1}{M - 1}
   \left( \overline{x^2} - \overline{x}^2 \right).
\label{eq_unc}
\end{eqnarray}
Again, in the case of the direct flux, the sum over $n$ is dropped and
$M = N$.

The uncertainty $\sigma_X$ is only an upper limit on the true
uncertainty since this quantity measures both the Monte Carlo noise
associated with running a finite number of photons {\em and} the
intrinsic variation in the quantity.  For example, the total scattered
flux from a nebula can be computed by directly summing every photon's
scattered weight.  The uncertainty, $\sigma_X$, in this quantity has a
contribution from the intrinsic variation of the scattered flux across
the nebula as well as the Monte Carlo noise.  An upper limit which
would be closer to the true uncertainty can be computed by computing
the point-by-point uncertainties in an image of the nebula.  The
point-by-point uncertainties can be calculated using
\begin{eqnarray}
\sigma_X(i,j)^2 & = & \frac{1}{M(i,j)(M(i,j)-1)} 
   \sum_\alpha \sum_n (x_n^\alpha -
   \overline{X(i,j)} )^2 \nonumber\\ & = & \frac{1}{M(i,j) - 1} \left(
   \overline{X(i,j)^2} - \overline{X(i,j)}^2 \right).
\end{eqnarray}
where $M(i,j)$ is the number of scatterings contributing to output
quantity in pixel $(i,j)$, the sum over $\alpha$ and $n$ is only done
for those photons which contribute to the output quantity in pixel
$(i,j)$, and $X(i,j)$ is calculated using eq.~\ref{eq_idij},
\ref{eq_is_image}, or \ref{eq_qij}.  In the case of the direct surface
brightness image, the sum over $n$ is dropped and $M(i,j)$ is the
number of photons contributing to pixel $(i,j)$.

The uncertainty in the quantity $X$ can then be calculated using
\begin{equation}
\sigma_X^2 = \sum_{(i,j)} \sigma_X{(i,j)}^2.
\label{eq_unc_image}
\end{equation}
Calculating the uncertainty with eq.~\ref{eq_unc_image} will result in
a lower value of the uncertainty because the intrinsic variation of
the quantity $X(i,j)$ across the nebula will not be included in the
uncertainty calculation.  This calculation of the uncertainty is still
only an upper limit as the uncertainty at a specific point in a nebula
will still include a contribution from photons having intrinsically
different weights due to having emerged from different depths in the
nebula.

\begin{figure*}[tbp]
\epsscale{2.1}
\plottwo{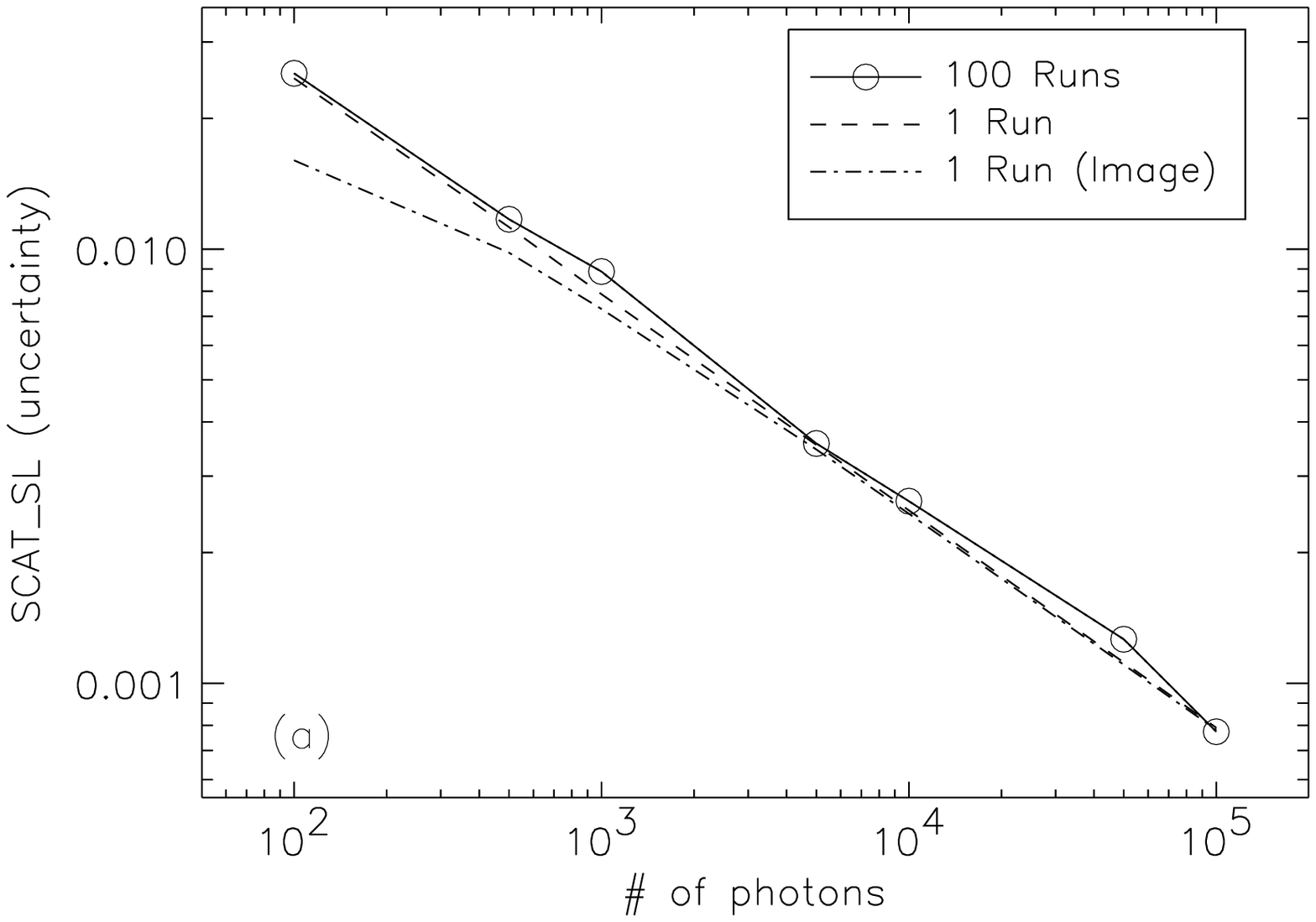}{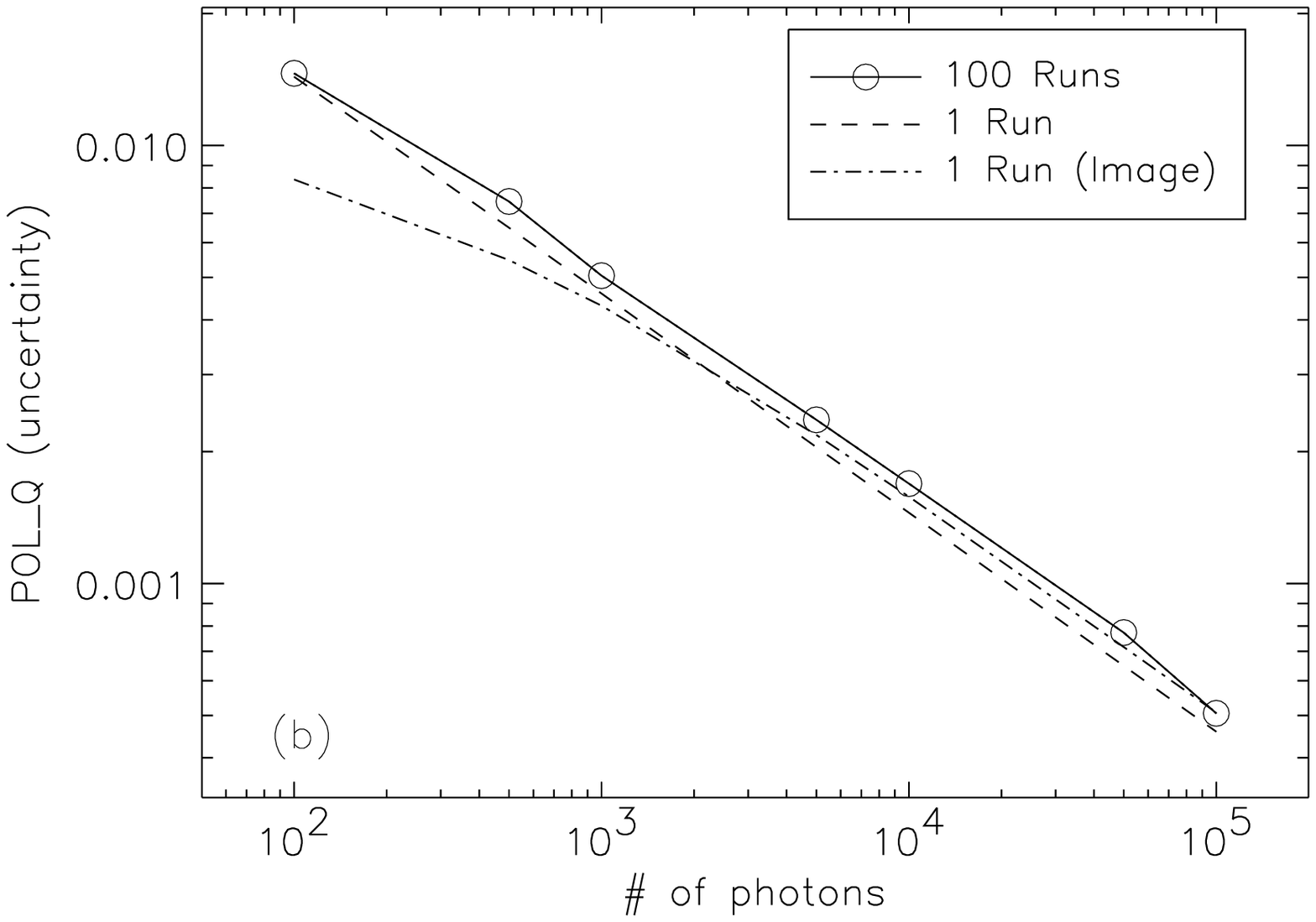}
\caption{The uncertainties in the scattered flux (a) and Q component of
the polarized flux (b) are plotted versus the number of photons run.
At each value of the number of photons, three values of the
uncertainties are plotted.  The solid line plots the uncertainty as
calculated from the variation in the 100 different runs, the dashed
line plots the average uncertainty as calculated for the total
quantity using eq.~\ref{eq_unc}, and the dot-dashed line plots the
average uncertainty as calculated using eq.~\ref{eq_unc_image}.  The
value, averaged over the 100 $10^5$ photon runs, of the scattered flux
(a) is 0.237 and the Q component of the polarized flux (b) is $-3.6
\times 10^{-5}$.  These numbers are for the case where $L = 1$
and $d = 1$.  Note that the Q component of the polarized flux is
equivalent to zero within the uncertainty, $(-3.6 \pm 5) \times
10^{-5}$, as expected for this spherically symmetric
model. \label{fig_test_unc}}  
\end{figure*}

We have tested this method of calculating uncertainties by running the
same model multiple times, but with a different random number seed
each time the model was run.  This allowed us to compute the true
uncertainty in an output quantity directly from the variation of the
quantity between model runs.  By comparing the true value of the
uncertainty with the values computed using eqs.~\ref{eq_unc} and
\ref{eq_unc_image}, we were able to evaluate the accuracy of our
simple method of estimating uncertainties in output quantities.  For
this testing, we chose to use a sphere with a homogeneous dust
distribution, a $\tau_V = 1$, V band Milky Way dust grain properties,
and a central illuminating star (see Fig.~\ref{fig_exam}a).  Other
optical depths give similar results.  We ran the model 100 times and
varied the total number of photons between $10^2$ and $10^5$.
Figure~\ref{fig_test_unc} displays the uncertainties in the scattered
flux and Q component of the polarized flux as a function of the number
of photons run.  The general trend is for the uncertainty calculated
using eq.~\ref{eq_unc} or eq.~\ref{eq_unc_image} to underestimate the
actual uncertainty by smaller amounts as the number of photons run
increases.  This is due to small number statistics, especially when
the uncertainty was calculated using eq.~\ref{eq_unc_image}.  For a
large number of photons (e.g., $10^5$), the uncertainty calculated
using eq.~\ref{eq_unc} or \ref{eq_unc_image} is a very good estimate
of the actual uncertainty.  The reason is that the majority of the
scattered flux and Q component of the polarized flux comes from the
central region of the nebula (see Fig.~\ref{fig_exam}a) and the
intrinsic variation of the scattered flux in the central region is
small.  This implies that the uncertainty calculated from
eq.~\ref{eq_unc} or \ref{eq_unc_image} is dominated by Monte Carlo
noise for this model.  There are model systems where this will not be
the case and, as a result, care must be taken calculating the
uncertainty using the method outlined above.

\subsection{Comparison with Other Models}

We tested the results of the DIRTY model against those produced by
Monte Carlo radiative transfer models which do not weight photons and
models which use the \citet{wit77} photon weighting.  These models
include ones which we have coded as well ones others have coded (J.\
Bjorkman 1999, private communication; K.\ Wood 1999, private
communication).  For computational reasons, these models are usually
restricted to spherically symmetric systems with smoothly varying
radial dust distributions.  Our two main test cases were for $\tau =
1$ and $\tau = 10$.  We adopted an albedo of 0.6 and a scattering
phase function asymmetry of 0.6.  In all cases, the \citet{yus84}
photon weighting method produced statistically similar results to the
other two weighting methods.  In addition, we computed the wavelength
dependence of the polarization for active galactic nucleus models
similar to those used by \citet{man96} and found qualitative agreement
with their results.  Quantitative agreement is more difficult to test
as we used a different dust grain model than \citet{man96}.

\begin{figure*}[tbp]
\epsscale{1.8}
\plottwo{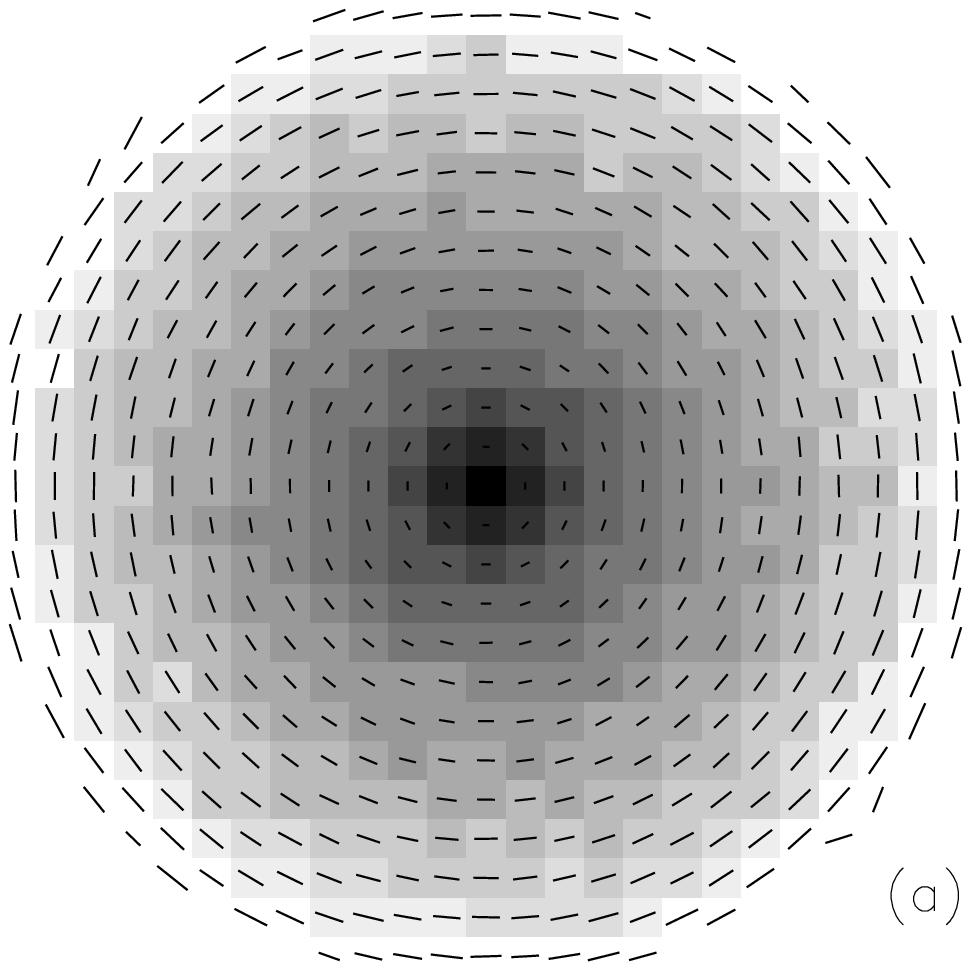}{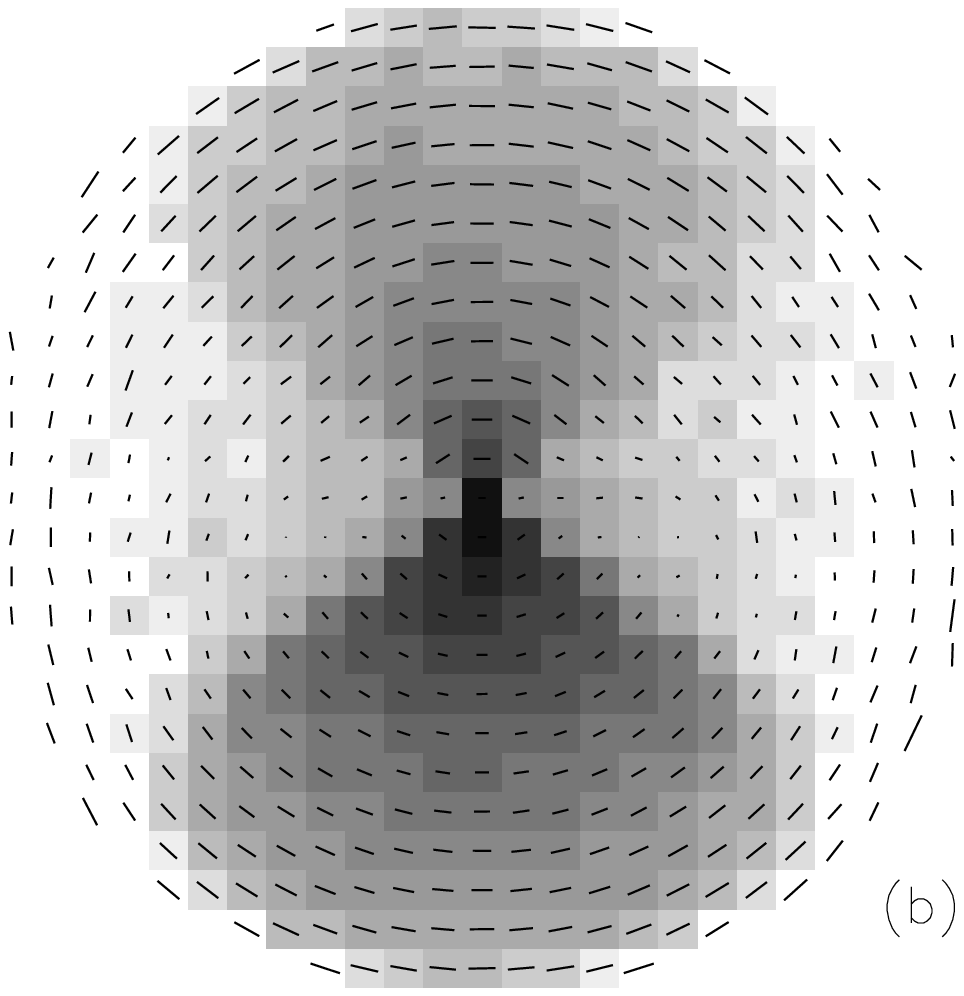}
\caption{Example images of the DIRTY model are shown for a spherical
system (a) and a conical system (b).  The spherical system is an
example of a reflection nebula with an embedded star.  The conical
system is an example of a bipolar nebula or a quasi-stellar object
with an opening angle of 45$\degr$ and an inclination of 30$\degr$
with the bottom lobe tilted toward the observer.  The grayscale
displays the distribution of scattered light and the line segments the
strength and orientation of the polarization.  The maximum length of
the line segments corresponds to a polarization of 48\% (a) and 56\%
(b).
\label{fig_exam}}   
\end{figure*}

\begin{figure*}[tbp]
\epsscale{1.5}
\plotone{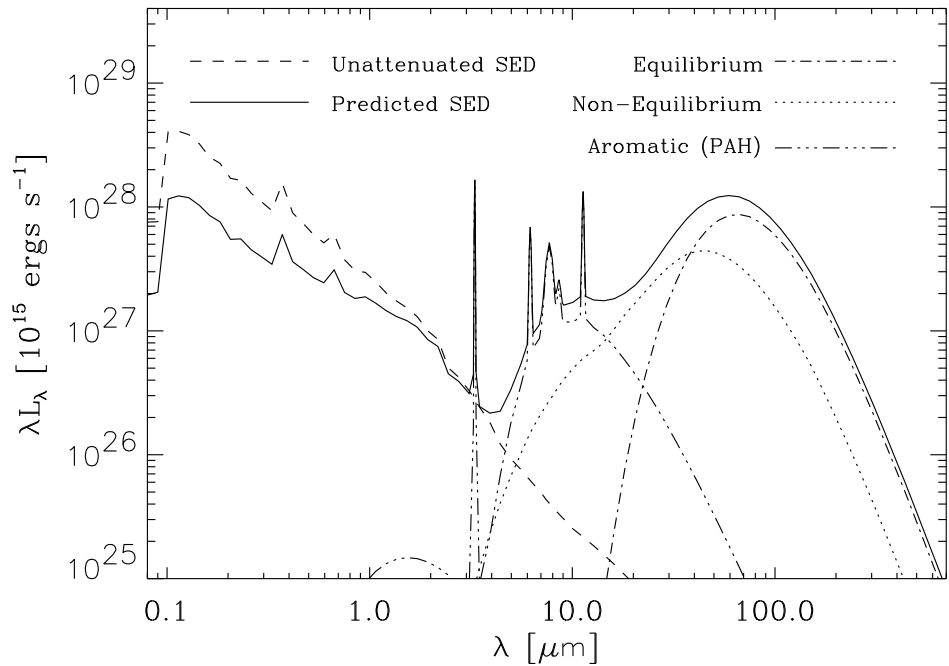}
\caption{The spectral energy distribution (SED) of a starburst stellar
population is plotted with and without dust.  The stellar population
is for constant star formation of 1.6 $M_{\sun}/year$ for a duration
of 100 million years.  The dust is Milky Way-like with a $\tau_V = 1$.
The dust distribution is the clumpy SHELL geometry (see \citet{mis00}
for additional details). \label{fig_exam_sed}}
\end{figure*}

Figure~\ref{fig_exam} illustrates the images produced by the DIRTY
model.  Figure~\ref{fig_exam}a shows how a spherical nebula with a
central illuminating star would look in the V band assuming Milky Way
type dust with a homogeneous distribution and a radial $\tau_V = 1$.
Figure~\ref{fig_exam}b shows how a biconical nebula or active galactic
nucleus inclined by an angle of 30$\degr$ would look in the V band
assuming Milky Way type dust with a homogeneous distribution and a
$\tau_V = 1$.  Figure~\ref{fig_exam_sed} displays the spectral energy
distribution before and after the inclusion of dust in a simple
starburst system.  This is a good illustration of how the dust
redistributes energy from the ultraviolet to the infrared.  In
addition, the three components (thermal equilibrium, thermal
non-equilibrium, and aromatic feature emissions) of the dust emission
spectrum are shown.  Starburst systems are investigated in detail in
\citet{mis00}.

\section{Summary and Discussion}

We have presented the DIRTY radiative transfer model in this paper and
a companion paper \citep{mis00}.  This model uses Monte Carlo
techniques to compute the radiative transfer of photons emitted from
arbitrary stellar distributions through arbitrary dust distributions.
The weighting scheme used in the DIRTY model results in the efficient
computation of the appearance of such systems from any viewing
direction.  The dust re-emission includes equilibrium thermal
emission, non-equilibrium thermal emission, and the aromatic feature
emission.  The dust emission is computed self-consistently with the
dust absorption and scattering using an iterative technique.  The
details of the dust emission are described by \citet{mis00}.

The DIRTY model can be used to study a wide range of astrophysical
objects due to its flexibility.  The general behavior of multi-phase
dust has been studied using the DIRTY model in reflection nebula
environments \citep{wit96} and galactic environments \citep{wit00}.
The very non-symmetric reflection nebula surrounding the R CrB star UW
Cen was successfully modeled using the DIRTY model \citep{cla99}.  The
optical depth of spiral galaxies was investigated using an early
version of the DIRTY model \citep{kuc98}.  The finding that the lack
of a 2175~\AA\ depression in the spectra of UV-selected starburst
galaxies was not due to radiative transfer effects but due to the dust
grain properties was accomplished with the DIRTY model \citep{gor97}.
The ultraviolet through near-IR spectral energy distribution of the
M33 nucleus was modeled using the DIRTY model and discovered to be a
post-starburst stellar population surrounded by a shell of Milky
Way-like dust \citep{gor99}.  The general behavior of the infrared
emission of starburst galaxies has been studied by \citet{mis00}.  The
full ultraviolet through far-infrared spectral energy distributions of
a handful of starburst galaxies was studied by \citet{mis00b}.  The
range of objects which the DIRTY model has already been applied to is
a good illustration of the usefulness of such a radiative transfer
model.

In addition to continuing to use the DIRTY model to explore the nature
of dust and its affects on a variety of astrophysical systems, we plan
to continue to improve the model itself.  One such improvement would
be to modify the current way the dust distribution is stored to allow
for grid points to be subdivided independently allowing for a larger
range of size scale to be efficiently modeled \citep{wol99}.  For
example, this would allow for modeling the physically small region
near the accretion disk in a QSO while simultaneously modeling the
large scattering regions in the jets of the QSO.  Another improvement
would be to include electron scattering which would improve the
application of the DIRTY model to the study of QSOs.  Finally, the
code could be parallelized to allow larger dust distribution grids to
be studied in shorter amounts of time.

As there are always more interesting astrophysical systems to be
studied than time in the day, anyone interested in using the DIRTY
model should contact the authors for possible collaboration.

\acknowledgements

We thank Jon Bjorkman and Kenny Wood for discussions on Monte Carlo
radiative transfer, especially on the subject of the proper weighting
of photons for particular lines-of-sight and how to compute the
uncertainties in output quantities.  We also thank the referee, Bruce
Draine, for his comments which substantially improved the clarity of
this paper.  Support for this work was provided by NASA through LTSA
Grants NAGW-3168 and NAG5-7933 and the ATP Grant NAG5-9203.  KAM
gratefully acknowledges financial support from the Louisiana Space
Consortium through NASA grant NGT5-40035.

\end{document}